\documentclass[twocolumn]{aastex62}
\usepackage{placeins}

\newcommand*{\Cratio}{^{12}\mathrm{C}/^{13}\mathrm{C}}
\newcommand*{\FeH}{[\mathrm{Fe/H}]}

\newcommand{\code}[1]{\texttt{#1}}
\newcommand*{\mesa}{\code{MESA}}

\begin{document}
\title{Carbon Isotope Ratios in M10 Giants}

\author{Z. G. Maas} 
\affil{Department of Astronomy, Indiana University, Bloomington, IN 47405, USA}

\author{J. M. Gerber}
\affil{Department of Astronomy, Indiana University, Bloomington, IN 47405, USA}

\author{Alex Deibel}
\affil{Department of Astronomy, Indiana University, Bloomington, IN 47405, USA}
\affil{The Joint Institute for Nuclear Astrophysics - Center for the Evolution of the Elements, Michigan State University, East Lansing, MI 48824, USA} 

\author{C. A. Pilachowski}
\affil{Department of Astronomy, Indiana University, Bloomington, IN 47405, USA}

\email{zmaas@indiana.edu}
\email{jgerb@indiana.edu}

\begin{abstract}

We measured carbon abundances and the $\Cratio$ ratio in 31 giant branch stars with previous CN and CH band measurements that span -2.33 $<$ M$_{\rm V}$ $<$ 0.18 in the globular cluster M10 (NGC 6254). Abundances were determined by comparing CO features at $\sim 2.3\, \mu \mathrm{m}$ and specifically the $^{13}$CO bandhead at $2.37\, \mu \mathrm{m}$, to synthetic spectra generated with MOOG. The observed spectra were obtained with GNIRS on Gemini North with a resolution of R $\approx 3500$. The carbon abundances derived from the IR spectra agree with previous [C/Fe] measurements found using CN and CH features at the near-UV/blue wavelength range. We found an average carbon isotope ratio of $\Cratio$ = 5.10$_{-0.17}^{+0.18}$ for first generation stars (CN-normal; 13 stars total) and $\Cratio$ = 4.84$_{-0.22}^{+0.27}$ for second generation stars (CN-enhanced; 15 stars). We therefore find no statistically significant difference in $\Cratio$ ratio between stars in either population for the observed magnitude range. Finally, we created models of the expected carbon, nitrogen, and $\Cratio$ surface abundance evolution on the red giant branch due to thermohaline mixing using the \mesa~stellar evolution code. The efficiency of the thermohaline mixing must be increased to a factor of $\approx 60$ to match [C/Fe] abundances, and by a factor of $\approx 666$ to match $\Cratio$ ratios. We could not simultaneously fit the evolution of both carbon and the $\Cratio$ ratio with models using the same thermohaline efficiency parameter. 

\end{abstract}


\section{Introduction}
\label{sec::intro}

Light element abundances in globular clusters (hereafter GCs), in particular from C to Ca, delineate mixing events during stellar evolution and reveal the presence of multiple populations of chemically distinct stars. Abundance patterns of proton-capture elements made during hydrostatic hydrogen burning $\textrm{--}$ such as C-N anti-correlations and Na-O anti-correlations $\textrm{--}$ constrain the possible formation scenarios for multiple populations in globular clusters \citep[see, e.g.,][and references therein]{gratton12}. Measurements of elements, such as C, N, and Li in giants explore the first dredge up and the non-canonical extra mixing. % (e.g. \citealt{angelou15}). 

The carbon isotope ratio $\Cratio$~is a particularly sensitive diagnostic of mixing in stars and evolutionary state. As a star evolves along the red giant branch (RGB) the CNO process will dominate in the hydrogen burning shell and the burning products are transported from the interior of the star to the surface \citep{iben67}. Stellar evolution theory predicts solar mass stars will undergo the first dredge up at the bottom of the red giant branch, which will pollute the surface with CNO process material and dilute Li \citep{iben65,charbonnel95}.

For some stars, however, the first dredge up is not the end of surface abundance changes. Low-mass red giant branch stars (0.5 $\leq$ $M_\odot$ $\leq$ 2.0) have been observed to undergo further extra mixing on the RGB after evolving past the luminosity function bump (LFB) \citep[e.g.,][and references therein]{suntzeff81,suntzeff91,gratton00,smith06,recio07,kirby15}. The LFB is a maximum in the luminosity function (LF) along the RGB that is a result of an evolutionary ``stutter" that occurs at that magnitude. Once the hydrogen burning shell of an RGB star burns out to the $\mu$-barrier (a gradient in mean molecular weight left behind at the inner most point of the convective envelope's penetration during the first dredge up), new H-rich material is introduced into the H-burning shell \citep{iben65}. This introduction of new material causes the star to burn hotter and bluer, and stalls its evolution until the star reaches equilibrium and continues up the RGB \citep{iben68,cassisi02}. The evolutionary pause creates a ``bump" in the luminosity function of globular clusters.

Many examples of this extra mixing in RGB stars are observed in the Milky Way. For example, field giants exhibit signs of extra mixing in their light element abundances. Surveys of metal-poor Population II giants find $\Cratio$~ ratios reaching the CNO equilibrium value of $\Cratio \sim 3.5$ \citep{sneden86}. Measurements of field giants in the metallicity range $-4 <\FeH<-1$ show isotope ratios between $\approx 3\textrm{--}10$ for giants beyond the RGB luminosity function bump with most between $\approx 6\textrm{--}10$  \citep{pilachowski97,charbonnel98,gratton00,keller01,spite06}. The $\Cratio$ ratio has also been measured as a function of evolutionary state and mass in open clusters \citep{smiljanic09,tautvaisiene16}; for example, $\Cratio$~ratios in RGB stars in the old open cluster NGC 6791 are between $\approx 6 \textrm{--} 11$, with the low values attributed to thermohaline mixing \citep{szigeti18}.

Similar to field giants, the $\Cratio$~ratio in GC giants also shows evidence for extra mixing. Since the evolutionary states of cluster stars are known, abundance studies in GCs can provide additional constraints on the extra-mixing process. For example, \citet{shetrone03} found the carbon isotope ratio is significantly lower than expected from the first dredge up for stars brighter than the luminosity function bump \citep{charbonnel98}. An additional complication in GCs is the dispersion of CNO surface abundances across multiple populations within a cluster. Differences in CNO abundances similar to those caused by extra mixing could arise from a second population of stars formed from gas ejected from those first population stars that have undergone proton-capture nucleosynthesis during their evolution, thereby changing their abundance of light elements. These second generation stars would then have depleted carbon, enhanced nitrogen, and could have lower $\Cratio$~ratios. For example, CN-enhanced stars in clusters have high nitrogen abundances and low carbon abundances showing evidence of an initial composition enriched with proton-capture material \citep[see, e.g.,][and references therein]{gratton12}.  

Differences in the carbon isotope ratio between CN-enhanced and CN-normal stars in the cluster M71 were originally found using measurements of the CN band at $\sim 8005\,\mbox{\AA}$. CN-normal stars had an average $\Cratio=8.3$ and CN-enhanced stars had an average of $\Cratio=6.0$ \citep{briley97}; this result was later confirmed with additional measurements using CO features at $2.3\, \mu \textrm{m}$ \citep{smith07}. These measurements were initially interpreted as evidence of primordial abundance variations because a correlation between band strength and additional mixing seemed unlikely \citep{briley97}. No correlation was found, however, between the carbon isotope ratio and their CN band strength in lower metallicity clusters M4 and NGC 6752 ($\FeH =-1.16$ and $\FeH =-1.54$, respectively compared to $\FeH =-0.78$ in M71) \citep{suntzeff91}. Carbon isotope ratios between $\approx 3\textrm{--}5$ were measured in nearly every sample star regardless of band strength; those values are lower than the ratios typically measured in metal poor field giants. A study from \citet{pavlenko03} of M71, M5, M13, and M3 giants found isotope ratios near the equilibrium value for all clusters. Metal poor clusters ($\FeH \lesssim -1.5$) have uniformly low isotope ratios between $\approx 3 \textrm{--}5$. These values are lower than typical carbon isotope ratios 6 $<$ $\Cratio$ $<$ 10 found in metal poor giants in the field \citep{gratton00,keller01}.

Multiple explanations have been proposed for the cause of this observed extra-mixing, as a non-canonical theory is needed to explain how CN(O)-cycle material is escaping the H-burning layer of the star and being brought to the surface by the convective envelope. A number of theories have been proposed, including rotational mixing \citep{sweigart79,chaname05,palacios06}, magnetic fields \citep{palmerini09,nordhaus08,busso07,hubbard80}, and internal gravity waves \citep{denissenkov00}. One of the most promising and well studied of these is thermohaline mixing or fingering convection, where mixing is caused by a molecular weight inversion created during $^{3}$He burning in the hydrogen burning shell \citep{eggleton06,charbonnel07}. The physical framework, multi-dimensional simulations, and one-dimensional approximations have been explored to determine the effects of thermohaline mixing on stars \citep{kippenhahn80,eggleton06,charbonnel07,denissenkov11,traxler11,wachlin11,brown13,henkel17}. Models created using different theoretical prescriptions have failed to match surface abundances unless the mixing efficiency is increased by $\sim 10 \textrm{--} 1000$ from predicted values \citep{angelou12,wachlin14,angelou15}.

In this study, we examine the $\Cratio$~isotope ratio in M10 (NGC 6254; $\FeH =-1.56$ \citealt{harris96} (2010 Edition)) giants with measured C and N abundances and constrain the mixing mechanism in these stars. In Section~\ref{sec:observations}, we discuss our observational methods and in Section~\ref{section:methodology} we present our measurements of the carbon isotope ratio in 31 giants. In Section~\ref{sec::discussion_abundances} we compare our carbon abundances to previous measurements in the literature and check for differences in the $\Cratio$~ in the multiple populations of M10. We compare the abundances to models in sections \ref{sec::models1} and \ref{sec::models2}. Finally, our conclusions are summarized in Section~\ref{sec::conclusion}.

	\begin{figure}[htp]
	\centering
	\includegraphics[trim=0cm 0cm 0cm 0cm, scale=.42]{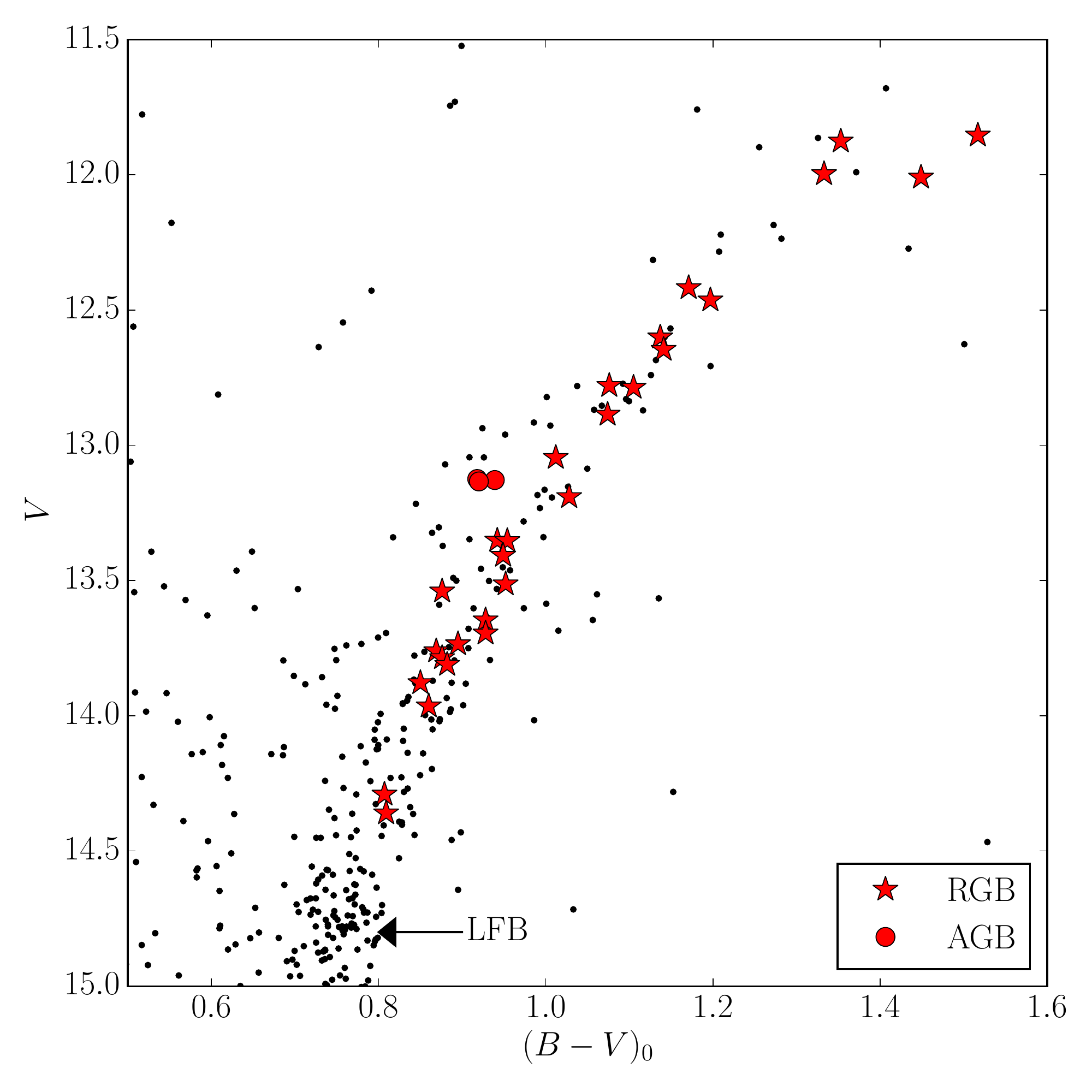}
	\caption{CMD of M10 using photometry from \citet{pollard05}. Our sample includes RGB stars plotted as red stars and AGB stars as red circles. The LFB of the cluster as measured by \citet{nataf13} is indicated on the CMD.}
	\end{figure}

\section{Observations and Data Reduction} \label{sec:observations}

We observed 31 stars along the giant branch in M10 that represent a range in absolute magnitude of -2.33 $<$ M$_{\rm V}$ $<$ 0.18. CO features become too weak to measure at temperatures hotter than $\sim 4900\, \mathrm{K}$ and surface gravities of $\mathrm{log}(g) > 2.3\,\mathrm{dex}$, for signal-to-noise ratios less than 200. Our sample comes from stars with measured C and N abundances from \citet{gerber18}. Since every star in our sample has known C and N abundances, we are able to measure C abundances from the CO features without having to assume [N/Fe]. The sample contains 14 CN-normal stars and 17 CN-enhanced stars allowing us to probe potential differences in the carbon isotope ratio present in multiple populations in globular clusters. 

\begin{deluxetable}{ c c c c c }
%\tabletypesize{10pt} 
%\rotate
\tablewidth{0pt} 
\tabletypesize{\footnotesize}
\tablecaption{Summary of GNIRS Observations \label{table::obslog}} 
 \tablehead{\colhead{ID\tablenotemark{a}} & \colhead{2MASS}  & \colhead{UT Date}& \colhead{M$_{\rm V}$} & \colhead{K$_{s}$\tablenotemark{b}}  \\
 \colhead{} & \colhead{Number} & \colhead{Observed} & \colhead{(Mag)} & \colhead{(Mag)}  }
\startdata
9  &  16571076-0404440   & 2017 Mar 6 & --2.32  &  7.621  \\
11 & 16570037-0406015 & 2017 Apr 9 &   --2.30  &  7.902  \\
14 & 16573237-0403007  & 2017 Feb 3 &  --2.18  &  8.060  \\
15 & 16570952-0407222 & 2017 Mar 21&   --2.17 &   7.837  \\
26 & 16570675-0403104  & 2017 Feb 4&   --1.76 &   8.782  \\
28 & 16571705-0413268  & 2017 Mar 21 & --1.72 &   8.765  \\
33 & 16573484-0407081  & 2017 Feb 4&   --1.58 &   8.949  \\
36 & 16571442-0411113 &  2017 Feb 2&   --1.53 &   8.998  \\
43 & 16570277-0410209 & 2017 Jan 24&   --1.40  &  9.285  \\
45 & 16570464-0406340 & 2016 Apr 9 &   --1.39   & 9.226  \\
53 & 16570550-0408364 & 2017 Mar 19 &  --1.29 &   9.391  \\
63 & 16571117-0403190 & 2017 Mar 8&    --1.13 &   9.666  \\
67\tablenotemark{c} & 16570782-0409370 & 2017 Apr 11&   --1.06  &  9.944  \\
68\tablenotemark{c} & 16563597-0405137 & 2017 Mar 23&   --1.05 &   9.853  \\
69\tablenotemark{c} & 16564584-0401260  & 2017 Mar 19&  --1.06 &   9.964  \\
73 & 16572587-0406175  & 2017 Feb 3&   --0.99 &   9.739  \\
83 & 16565465-0408554  & 2017 Apr 13&  --0.83 &   10.089 \\
84 & 16565750-0406502  & 2017 Apr 13&  --0.83  &  10.070 \\
88 & 16565433-0405359  & 2017 Apr 11&  --0.77 &   10.150 \\
97 & 16571608-0404329  & 2017 Apr 11&  --0.67 &   10.204 \\
102 & 16571668-0407102  & 2017 Apr 13& --0.64 &   10.369 \\
115 & 16571126-0407476  & 2017 Apr 13& --0.53 &   10.357 \\
119 & 16570116-0409350  & 2017 Apr 13& --0.49 &   10.483 \\
122 & 16565359-0408144  & 2017 Apr 13& --0.45 &   10.571 \\
127 & 16565514-0412188  & 2017 May 7&  --0.42 &   10.617 \\
131 & 16565831-0408052  & 2017 Apr 28& --0.39 &   10.651 \\
137 & 16565101-0410349  & 2017 Apr 13& --0.37 &   10.645 \\
143 & 16570249-0402413  & 2017 May 12& --0.30 &   10.881 \\
158 & 16570774-0404433  & 2017 May 12& --0.22 &   10.870 \\
213 & 16570581-0405159  & 2017 May 29& 0.11 &     11.304 \\
222 & 16570363-0412481  & 2017 May 23& 0.18 &     11.344 \\
\enddata
\tablenotetext{a}{ID From \citet{pollard05}}
\tablenotetext{b}{K$_{s}$ magnitudes from 2MASS \citep{skrutskie06}}
\tablenotetext{c}{AGB Star}
\end{deluxetable}

Observations were obtained with GNIRS on Gemini North as apart of GN-2017A-Q-70. We used the short camera, the 0.3" slit, and the $110.5\, \mathrm{lines \ mm^{-1}}$ grating to achieve a resolution of $R\approx 3500$ in the K band. We observed the wavelength range between $22800\,\mbox{\AA}\textrm{--}24000\,\mbox{\AA}$ to measure $^{13}$C$^{16}$O features at $2.34\, {\mu}\mathrm{m}$ and $2.37\, {\mu}\mathrm{m}$. Targets were observed by nodding along the slit in an 'abba' pattern to remove sky contamination. The star HIP~82162 was observed as a telluric standard for each night of observation. Exposure times sufficient to achieve S/N ratios of 100 per pixel for our target stars were calculated for non-ideal observing conditions. Observations obtained using GNIRS ranged in exposure time from 20s (for four images) for ID 9 to 240s (for eight images) for ID 222.

Data reduction was carried out using the IRAF\footnote{IRAF is distributed by the National Optical Astronomy Observatory, which is operated by the Association of Universities for Research in Astronomy, Inc., under cooperative agreement with the National Science Foundation.} software suite; specifically, the Gemini IRAF package. The task \texttt{nsprepare} was used to subtract an offset from the images and prepare the headers for further data reduction. The object files were then subtracted from each other to test for pattern noise that sometimes occurs in the IR detectors on Gemini. The python script \texttt{cleanir.py}\footnote{Obtained from \url{https://www.gemini.edu/sciops/instruments/niri/data-format-and-reduction/cleanir}} was used to remove the pattern noise when detected in our images. The flat field images were combined and the object images were trimmed, flatfield corrected, and subtracted from one another (for sky-subtraction and bias removal) using the task \texttt{nsreduce}. The object images were extracted using the task \texttt{nsextract}. 

The telluric lines were removed by dividing the object spectra with the spectrum of HIP~82162. HIP~82162 is an A1VI star with no known binary and is in close position on the sky to the cluster M10 to match airmass between the telluric standard and object observations. The wavelength solution was determined using the telluric lines in the spectrum of HIP~82162 with line identifications from \citet{lord92}. The final spectra were average combined and normalized.

	\begin{figure}[htp]
	\centering
	\includegraphics[trim=.6cm 0cm 0cm 0cm, scale=.85]{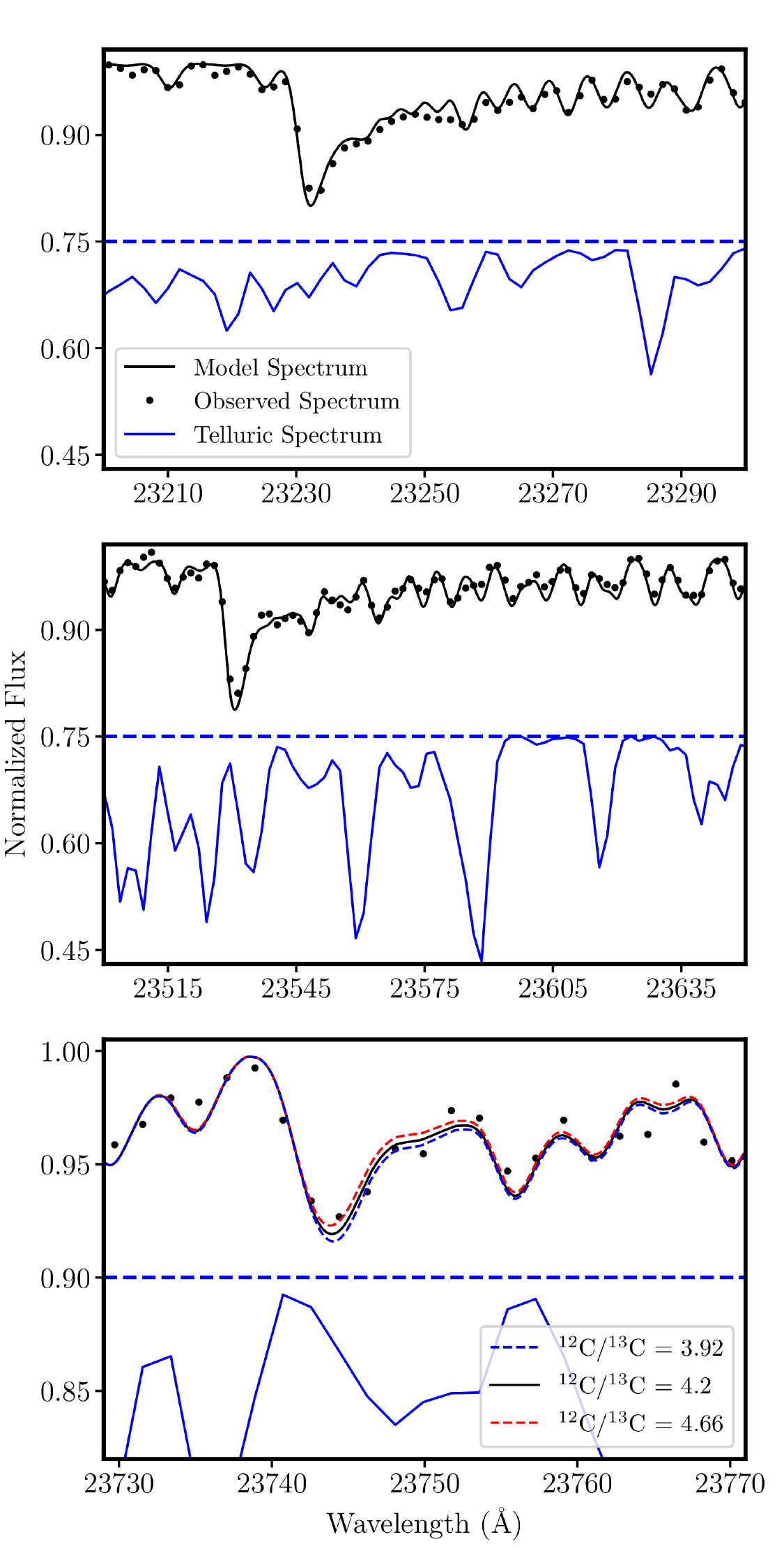}
	\caption{Spectra for the star ID 83 in the regions used to determine abundances.  Black points are the observed spectrum, the black line is the best model fit, the blue line represents the telluric spectrum (offset from 1 by a constant value). Dashed blue lines represent the continuum for the offset telluric spectrum. The top and middle panel show the  $^{12}$CO lines used to determine carbon abundances. The bottom panel shows the fit to the $^{13}$CO bandhead with the red and blue dashed lines representing the lower and upper 1$\sigma$ uncertainties respectively. \label{fig:spectra_id83}  }
	\end{figure}

\section{Measuring the Carbon Isotope Ratio}
\label{section:methodology}
\subsection{Carbon Abundance and Isotopic Ratio Derivation}
\label{sec::Carbon Abundance and Isotopic Ratio Derivation}

Carbon isotope ratios in our stars were derived by comparing the normalized spectrum for each star to synthetic spectra generated with different isotope ratios. The synthetic spectra were created using MOOG \citep{sneden73} and MARCS atmospheric models \citep{gustafsson08}. The line-list was compiled using CO transitions from \citet{goorvitch94} and atomic transitions from the Kurucz database\footnote{\texttt{http://kurucz.harvard.edu/atoms.html}}. The effective temperatures and surface gravities for our sample were derived by \citet{gerber18} using color relations from \citet{alonso99,alonso01}. A metallicity of $\FeH = -1.56$ for M10 was adopted from the Harris Catalogue \citealt{harris96} (2010 edition). The full list of atmospheric parameters can be found in Table \ref{table::Abundances}. A microturbulence ($\xi$) of $2.0\, \mathrm{km \ s^{-1}}$ was adopted as a reasonable approximation for stars on the red giant branch, which matches what was adopted by \citet{gerber18}. To compare, \citet{carretta09b} measured an average microturbulence of $1.84 \pm 0.24$ (st.dev.) for fourteen stars in M10 with a temperature range of $T = 4381\,\mathrm{K}\textrm{--} 4737\,\mathrm{K}$. One star, 2MASS J16571278-0403156, had an anomalous $\xi = 1.11 \mathrm{\ km \ s^{-1}}$ and if removed from the sample, the average microturbulence becomes $1.89 \pm 0.22$, consistent with the value chosen for our stars. 

The MARCS atmospheric models are spherical and 1-D. Also, MOOG assumes LTE when creating the synthetic spectra. The assumption of LTE is not expected to cause systematic errors when deriving abundances since CO lines are not impacted by NLTE effects \citep{ayres89,schweitzer00}. The use of 1-D atmospheric models may however not be appropriate for CO lines. An analysis of 3D models versus 1D models found differences in CO line formation \citep{dobrovolskas13} for giant stars at $T \sim 5000\,\mathrm{K}$, $\mathrm{log}(g)=2.5$, and at [M/H] from 0 to --3 in steps of 1 dex. The abundance correction for CO lines at $\FeH = -2$ is $\sim -0.5\, \mathrm{dex}$ while the correction at [M/H] = --1 is near zero for lines with excitation potentials between 0 and 1 eV at $1600\,\mathrm{nm}$. The effects on other molecular lines are smaller than on CO. We empirically test our carbon abundances from CO lines by comparing them to [C/Fe] derivations from CN and CH lines from \citet{gerber18} in section \ref{subsec::carbon_comparison}. 

Synthetic spectra were generated using the initial C and N abundances of \citet{gerber18} and oxygen abundances from \citet{carretta09a,carretta09b} when available. Average oxygen abundances were adopted for stars without known oxygen abundances: [O/Fe] = $0.39\, \mathrm{dex}$ for CN-normal stars and [O/Fe] = $0.11\, \mathrm{dex}$ for CN-enhanced stars as in \citet{gerber18}. 

A grid of synthetic spectra with carbon abundances in steps of $0.01\, \mathrm{dex}$ was synthesized and compared to the $23300\,\mbox{\AA}\textrm{--}23400\,\mbox{\AA}$ and $23500\,\mbox{\AA}\textrm{--}23650\,\mbox{\AA}$ regions of the observed spectra to find the best fitting carbon abundance. These regions were chosen to derive [C/Fe] ratios because they do not contain strong $^{13}$CO features or blanketing by the strongest telluric features (at wavelengths $>$23800 $\mbox{\AA}$). The carbon abundance that minimized the $\chi^{2}$ between the synthetic spectrum and the fit was found for each region and averaged together to determine the final carbon abundance for each star, listed in Table~\ref{table::Abundances}. The regions used to determine the carbon abundance are shown in Figure \ref{fig:spectra_id83}. The smoothing factor and velocity shifts for each star were determined by eye for each spectrum initially. A signal to noise ratio of 100 was assumed for each star. 

Carbon abundance derivations were tested by fitting MOOG synthetic spectra with Markov chain Monte Carlo (MCMC) methods. The smoothing parameters, carbon abundance, and an additional factor for the signal to noise were assumed as free parameters. The log likelihood equation took the form:

\begin{equation}\label{equation:loglikelihood}
\mathrm{log}(L) = -0.5 \sum \frac{D - \mathrm{synth}(S,\mathrm{A(C)})}{(\sigma*f)^2.}  + \mathrm{log}\left(\frac{1}{(\sigma*f)^2} \right) \, ,
\end{equation}

%In Equation~\ref{equation:loglikelihood}, 
\noindent where $D$ represents the observed data, synth is the synthetic spectrum created by MOOG with a Gaussian smoothing factor with a full width half max of $S$ and carbon abundance of A(C), and $f$ represents deviations from the expected signal to noise of 100 ($\sigma$ was assumed to be 0.01). The \texttt{emcee} package \citep{foreman13} was used to find the optimal abundances using the star ID 83 as an example. For the region between $23300\,\mbox{\AA}\textrm{--}23400\,\mbox{\AA}$ the MCMC method found a [C/Fe] = --0.33 $\pm$ 0.02, $S = 3.89 \pm 0.16 \, \mbox{\AA}$, and $f = 0.89 \pm 0.08$; and for $23500\,\mbox{\AA}\textrm{--}23650\,\mbox{\AA}$, [C/Fe] = --0.42 $\pm$ 0.02, $S = 3.52^{+0.35}_{-0.36}$, and $f = 1.16^{+0.11}_{-0.10}$. The uncertainties are the 14$\%$ and 86 $\%$ values in the distribution of results. The average carbon abundance for these two regions is --0.38, which is exactly the result found from the previous method. The signal to noise ratio was also approximately consistent with 100 for each region.

The carbon isotope ratios were also determined by creating a grid of synthetic spectra in steps of 0.1 over the range $2.0 <$  $\Cratio$ $< 10$. The synthetic spectra were fit to the $^{13}$CO band-head at $\sim 2.37\,\mu\mathrm{m}$ as shown in Figure \ref{fig:spectra_id83}. Isotope ratios were not determined from the $2.34\,\mu \mathrm{m}$ bandhead because this spectral region has stronger telluric absorption features blended with the CO lines, the continuum is more difficult to define, and the features at $2.34\, \mu \mathrm{m}$ are less sensitive to the $^{13}$C abundance than the $2.37\,\mu\mathrm{m}$ feature. The best fit $\Cratio$~ ratio is given in Table \ref{table::Abundances}.

\begin{deluxetable}{lllllllll}
%\tabletypesize{10pt} 
%\rotate
\tablewidth{0pt} 
\tabletypesize{\scriptsize}
\tablecaption{Atmospheric Parameters and Carbon Isotope Ratios \label{table::Abundances}} 
 \tablehead{\colhead{ID}  & \colhead{T$_{\rm eff}$} & \colhead{log g} &  \colhead{[C/Fe]\tablenotemark{a}} & \colhead{[N/Fe]\tablenotemark{a}} & \colhead{[O/Fe]\tablenotemark{b}} & \colhead{[C/Fe]\tablenotemark{c}} &  \colhead{$^{12}$C/$^{13}$C} }
\startdata
  9  &   3964& 0.68 &  -1.00	&	1.59	& 	-0.11     & --0.88 $\pm$ 0.14 &   4.2$^{+0.47}_{-0.30}$\\  
  11 &   4098& 0.81 &-0.56	&	0.77	& 	0.23  & --0.52 $\pm$ 0.16  &  4.1$^{+0.57}_{-0.35}$  \\  
  14 &   4120& 0.88 &-0.72	&	1.22	& 	\nodata & --0.73 $\pm$ 0.15  &  5.1$^{+0.37}_{-0.39}$  \\
  15 &   3994& 0.77 &-1.13	&	1.77	& 	-0.26 & --0.98 $\pm$ 0.14  &  4.1$^{+0.30}_{-0.30}$  \\  
  26 &   4310& 1.19 &-0.78	&	1.62	& 	\nodata & --0.94 $\pm$ 0.20 &  4.3$^{+0.64}_{-0.61}$  \\
  28 &   4267& 1.17 &-0.72	&	1.7	& 	\nodata     & --0.86 $\pm$ 0.21  &  4.6$^{+0.44}_{-0.33}$  \\
  33 &   4299& 1.25 &-0.36	&	0.82	& 	\nodata & --0.37 $\pm$ 0.15  &  5.3$^{+0.49}_{-0.68}$  \\
  36 &   4301& 1.27 &-0.36	&	0.65	& 	\nodata & --0.49 $\pm$ 0.30  &  5.5$^{+0.46}_{-0.47}$  \\
  43 &   4412& 1.40 &-0.4	&	0.77	& 	\nodata & --0.50 $\pm$ 0.16     &  3.5$^{+0.30}_{-0.24}$ \\
  45 &   4364& 1.37 &-0.73	&	1.15	& 	-0.07 & --0.61 $\pm$ 0.17  &  4.5$^{+0.45}_{-0.39}$  \\  
  53 &   4413& 1.44 &-0.39	&	1.09	& 	0.32 & --0.54 $\pm$ 0.16  &  4.2$^{+0.33}_{-0.30}$  \\  
  63 &   4505& 1.56 &-0.82	&	1.48	& 	-0.34 & --0.53 $\pm$ 0.25  &  3.8$^{+0.63}_{-0.42}$  \\  
  67 &   4677& 1.69 &-0.46	&	0.91	& 	0.53 & --0.45 $\pm$ 0.18  &  4.3$^{+0.61}_{-0.47}$  \\  
  68 &   4584& 1.64 &-0.64	&	1.42	& 	0.17 & --0.52 $\pm$ 0.23  &  4.0$^{+0.65}_{-0.44}$  \\  
  69 &   4688& 1.70 &-0.44	&	1.27	& 	0.36 & --0.36 $\pm$ 0.22  &  5.2$^{+0.82}_{-0.53}$  \\  
  73 &   4447& 1.58 &-0.2	&	0.8	& 	0.61      & --0.36 $\pm$ 0.14      &  5.0$^{+0.51}_{-0.37}$  \\  
  83 &   4595& 1.74 &-0.27	&	0.69	& 	0.57 & --0.38 $\pm$ 0.17  &  4.7$^{+0.46}_{-0.35}$  \\  
  84 &   4576& 1.73 &-0.39	&	1.19	& 	0.40 & --0.39 $\pm$ 0.18  &  4.6$^{+0.47}_{-0.35}$  \\  
  88 &   4601& 1.76 &-0.53	&	1.22	& 	0.13 & --0.46 $\pm$ 0.21 &  3.6$^{+0.42}_{-0.32}$  \\  
  97 &   4564& 1.78 &-0.36	&	0.74	& 	0.43 & --0.29 $\pm$ 0.17 &  3.8$^{+0.45}_{-0.24}$  \\  
  102 & 4687& 1.87 &-0.54	&	0.84	& 	\nodata & --0.50 $\pm$ 0.21  &  3.9$^{+0.62}_{-0.42}$  \\
  115 & 4570& 1.84 &-0.36	&	0.75	& 	\nodata & --0.31 $\pm$ 0.18   &  5.8$^{+0.60}_{-0.35}$  \\
  119 & 4646& 1.90 &-0.22	&	0.59	& 	\nodata & --0.13 $\pm$ 0.19 &  4.9$^{+0.60}_{-0.47}$  \\
  122 & 4694& 1.95 &-0.42	&	1.05	& 	\nodata & --0.33 $\pm$ 0.21  &  5.0$^{+0.84}_{-0.54}$  \\
  127 & 4715& 1.97 &-0.52	&	1.34	& 	\nodata & --0.26 $\pm$ 0.28  &  3.8$^{+1.11}_{-0.57}$  \\  
  131 & 4724& 1.98 &-0.34	&	0.56	& 	0.10 & --0.03 $\pm$ 0.18  &  5.1$^{+0.63}_{-0.50}$  \\  
  137 & 4692& 1.98 &-0.35	&	1.04	& 	\nodata & --0.31 $\pm$ 0.23  &  3.8$^{+0.53}_{-0.32}$  \\
  143 & 4870& 2.10 &-0.12	&	0.74	& 	0.45 & --0.09 $\pm$ 0.21  &  6.1$^{+0.82}_{-0.65}$  \\  
  158 & 4767& 2.08 &-0.31	&	1.2	& 	0.09     &  0.08 $\pm$ 0.28  &  5.9$^{+1.02}_{-0.65}$  \\  
  213 & 4883& 2.27 &-0.14	&	0.58	& 	0.47 &  0.02 $\pm$ 0.24  &  6.5$^{+1.03}_{-0.62}$  \\  
  222 & 4850& 2.28 &-0.4	&	1.34	& 	-0.01    & --0.04 $\pm$ 0.30  &  5.2$^{+2.30}_{-0.84}$  \\  
\enddata
\tablenotetext{a}{[C/Fe] From \citet{gerber18}}
\tablenotetext{b}{[O/Fe] From \citet{carretta09a,carretta09b}}
\tablenotetext{c}{[C/Fe] From This Work}
\end{deluxetable}

\subsection{Uncertainties}

Uncertainties on the atmospheric parameters are $\delta T = \pm 100\, \mathrm{K}$, $\delta \mathrm{log}(g) = \pm 0.20$, and $\delta$ microturbulence = $\pm 0.3 \, \mathrm{km \ s^{-1}}$. The uncertainties on the temperature and $\mathrm{log}(g)$ parameters were estimated from the scatter in the color-$T_{\rm eff}$-$\mathrm{log}(g)$ relation of \citet{alonso99} and consistent with the results of \citet{gerber18}. The uncertainty on the microturbulence was adopted from the scatter in $\xi$ from \citet{carretta09b}, discussed in section \ref{sec::Carbon Abundance and Isotopic Ratio Derivation}. Synthetic spectra were created using model atmospheres for one sigma variation of each atmospheric parameter. For example, for each star synthetic spectra were created with an effective temperature +100 K higher with all other atmospheric parameters held constant for a grid of carbon abundances and isotope ratios. The same $\chi^{2}$ minimization technique used to determine the best fit carbon abundance and isotope ratio was used with the new atmospheric parameters. The difference between the abundances from the derived atmospheric parameters at the one-sigma level was taken as the uncertainty on the abundance measurement. The average uncertainties on the carbon abundance and carbon isotope ratio for each atmospheric parameter are shown in Table~\ref{table::uncertainty}.

\begin{deluxetable}{ c c c c } 
%\tabletypesize{10pt} 
%\rotate
\tablewidth{0pt} 
\tabletypesize{\footnotesize}
\tablecaption{Average Uncertainties from Atmospheric Parameters \label{table::uncertainty}} 
 \tablehead{\colhead{Atmospheric} & \colhead{$\delta$A(C)}  & \colhead{$\delta$($^{12}$C/$^{13}$C)} & \colhead{$\delta$($^{12}$C/$^{13}$C)}   \\
 \colhead{Parameter} & \colhead{(dex)} & \colhead{ + 1$\sigma$} & \colhead{-- 1$\sigma$} }
\startdata
$T_{\rm eff}$ ($\pm$ 100 K)  &  0.18 & 0.11 & 0.08  \\
log($g$) ($\pm$ 0.2 dex) & 0.05 & 0.10 & 0.09  \\
$\xi$ ($\pm$ 0.30 km s$^{-1}$) & 0.03 & 0.25 & 0.16  \\
A(O) $\pm$ 0.07 & 0.03 & 0.1 & 0.1 \\
A(N) $\pm$ 0.25 & 0.01 & 0.0 & 0.0 \\
\enddata
\end{deluxetable}

In addition to changes to the models, we also computed the uncertainty in our derived carbon abundances and $\Cratio$ from variations in the nitrogen and oxygen abundance. Differences in any CNO abundance will affect the molecular equilibrium and therefore change the strength of molecular absorption features including CNO elements, such as CO. To test the magnitude of this effect on abundance determinations we calculated abundances while independently changing the O and N abundances for six representative stars in our sample. Pairs of CN-enhanced and CN-normal stars were chosen at the coolest, average, and warmest portion of our temperature range to discern if changes in line strength from changing the CNO abundances correlate with atmospheric parameters. The six stars chosen were ID 9, 11, 63, 83, 143, and 158. The average uncertainty on the oxygen abundances from \citet{carretta09b} is 0.07 dex and the typical uncertainty on the nitrogen abundances from \citet{gerber18} is 0.25 dex. Abundances were determined with the abundances of N and O varied by 1$\sigma$ independently. Varying the oxygen abundance resulted in an average uncertainty of 0.03 $\pm$ 0.02 dex (st.dev.) and contributed an uncertainty of 0.10 to the carbon isotope ratio (see Table \ref{table::uncertainty}). Varying the nitrogen abundance effected our derived [C/Fe] and $\Cratio$ ratios even less. 

The uncertainty on the carbon isotope ratio and carbon abundances based on the synthetic spectrum fit was also calculated and added in quadrature with the uncertainties from the atmospheric parameters and abundances. A signal to noise estimate of 100 was adopted for the final spectrum of each object. The synthetic spectra were re-fit to the data over multiple iterations to estimate the uncertainty from the data signal-to-noise ratio. For each iteration, each data point was represented by a random number generated from a Gaussian distribution with the mean equal to the data point value and a standard deviation derived from the signal-to-noise ratio of the data. The 5$\%$ and 95$\%$ value of carbon abundance and $^{12}$C/$^{13}$C ratio from the posterior distribution of the Monte Carlo simulation were accepted as the 2$\sigma$ uncertainty on the fit. The typical 1$\sigma$ uncertainty on the carbon abundance was small, on average the error was $\sim 0.05\, \mathrm{dex}$. However, the uncertainty on the fit was the dominant uncertainty term on the carbon isotope ratio, especially for warmer stars with stronger surface gravities and weak CO features. The final total uncertainties are given in Table~\ref{table::Abundances}.

\section{Discussion}
\label{sec::discussion_abundances}
\subsection{Carbon Abundance Comparisons to G18}
\label{subsec::carbon_comparison}

\begin{figure}[htp] %[t!]
	\centering
	\includegraphics[trim=1cm 0cm 0cm 0cm, scale=.72]{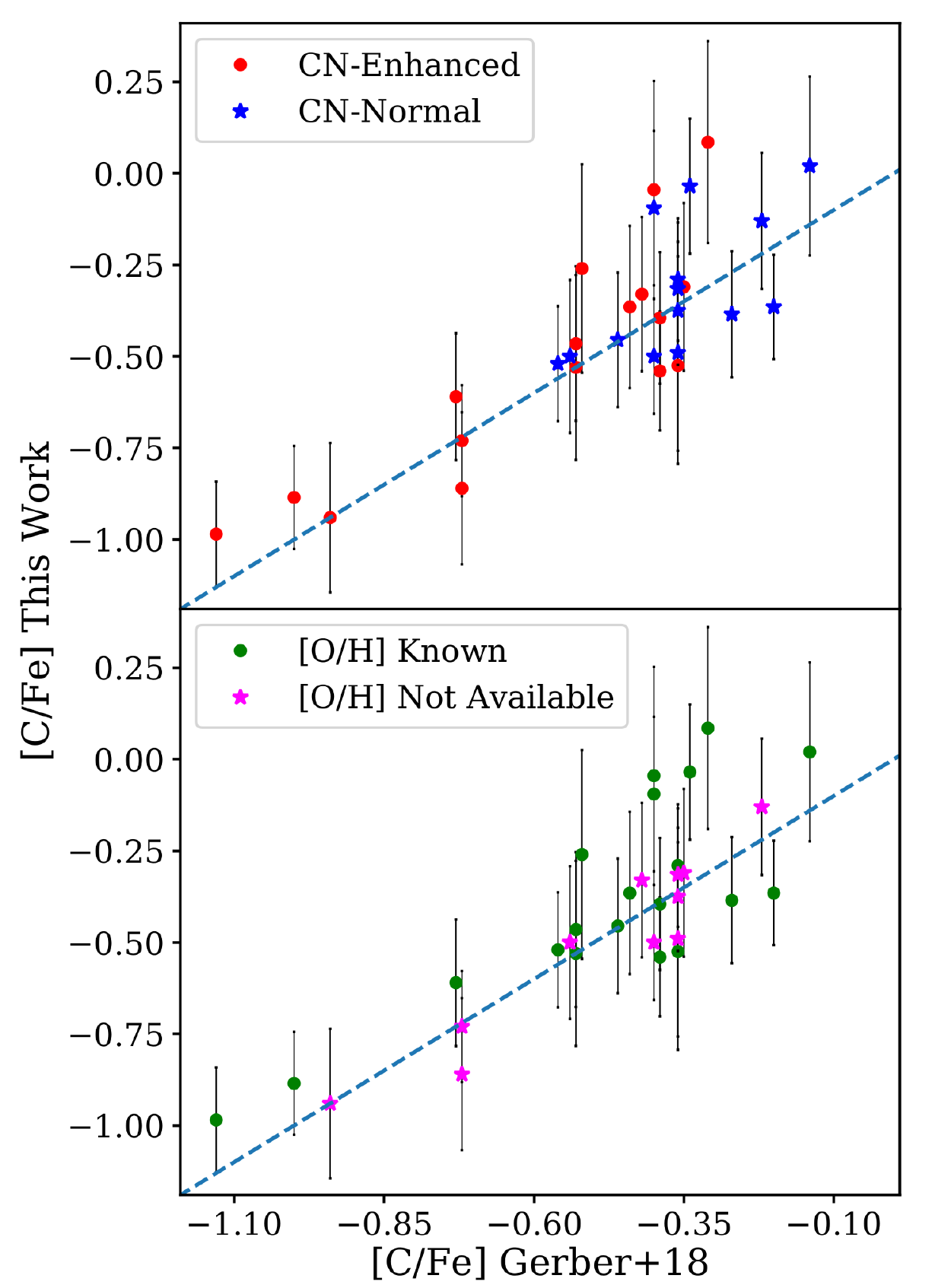}
	\caption{Two plots compare carbon abundances from IR spectra to abundances derived in \citet{gerber18}. The dashed line in both plots represents a line with a slope of one. The top panel shows red circles as CN-enhanced stars and blue stars as CN-normal stars. The bottom panel shows stars with known oxygen abundances as green circles and red giants with unknown oxygen abundances as pink stars.  \label{fig:carbon_ir_visible}  }
	\end{figure}

The carbon abundances derived from CO lines at $23200 \textrm{--} 23700\, \mbox{\AA}$ can be compared to abundances measured using the CN band at $\sim 3800 \, \mbox{\AA}$ and CH band at $\sim 4300\, \mbox{\AA}$ from \citet{gerber18}. Both [C/Fe] determinations are in reference to the solar abundances from \citet{asplund09}. Figure~\ref{fig:carbon_ir_visible} shows how the two sets of abundances compare to a line with a one-to-one slope. We also identify stars by CN strength in the top panel of Fig. \ref{fig:carbon_ir_visible} and by known O abundance in the bottom panel. These identifications show that there are no offsets between C abundances based on either of these two factors. We find the average difference between our abundances is $<$[C/Fe]$_{\mathrm{CN}}$ - [C/Fe]$_{\mathrm{CO}}$ $>$ = --0.06 $\pm$ 0.16 (st.dev.). To test further the consistency between the two determinations a linear fit using all our stars was performed using the MCMC code \texttt{emcee} \citep{foreman13}. The derived slope is 0.88$^{+0.15}_{-0.12}$ where the quoted uncertainties are the 14$\%$ and 86$\%$ percentile values in the slope distribution. We therefore conclude the carbon abundances from both studies are compatible. 

We note a small systematic offset for some of our [C/Fe] ratios when compared to \citet{gerber18}. Specifically, there is some preferential scatter towards higher [C/Fe] abundances that primarily exists with our warmest stars (4 of the 5). The T$_{\mathrm{eff}}$ vs color relation of \citet{alonso99} is likely not responsible as this relationship is accurate within the color range of 2 $<$ V - K  $<$ 4.6 and our warmest star (ID 213) has a V - K of 3.03. Additionally, measurements errors are likely not the cause as the $^{12}$CO bandheads are still strong enough to measure in the warmer stars. The systematic scatter may be due from random noise fluctuations and the offset for high temperature stars may be caused by small sample statistics. Second, the offset may be due to 1D vs. 3D model effects on the order of 0.1 to 0.4 dex. The systematic effects of using 1D models to derive abundances gain significance with T$_{\mathrm{eff}}$ and may also cause the offset between the \citet{gerber18} [C/Fe] ratios and our results \citep{dobrovolskas13,kucinskas13}. Additional measurements would be needed to distinguish between these hypotheses.

\citet{gerber18} measured carbon abundances by fitting CN and CH features, at 3883 $\mbox{\AA}$ and 4300 $\mbox{\AA}$ respectively using synthetic spectra generated with the Synthetic Spectrum Generator (SSG). The consistency also suggests 3D effects are similar for [C/Fe] measurements between CO measurements at lower metallicities and CH and CN features. The CH and CN features are expected to be minimally affected by 3D effects at both [Fe/H] = --1 and --2, while CO is predicted to have large corrections at an $\FeH = -2$ for stars at the base of the red giant branch \citep{dobrovolskas13}. The 3D effects are non-linear with metallicity and at the M10 iron abundance of $\FeH = -1.56$ the differences calculated between 1D and 3D model atmospheres have course grid spacing in [Fe/H] \citet{dobrovolskas13}. In other stars, the effects of 1D versus 3D models when deriving abundances with CO lines are significant and increase with increasing T$_{\mathrm{eff}}$ and log(g). For example, accounting for 3D effects is necessary when measuring abundances from CO lines in the Sun \citep{scott06}. Finally, C measurements in Arcturus with CO lines \citep{pavlenko10} ($\mathrm{A(C)} = 8.22 \pm 0.1\,\mathrm{dex}$) have agreed with measurements from C I lines ($\mathrm{A(C)} = 8.34 \pm 0.07\, \mathrm{dex}$) \citep{ramirez11}, showing the 1D approximation can yield accurate abundances with CO lines in red giants of $\FeH \sim -0.5$.

\subsection{Carbon Isotope Ratios in CN-Normal and CN-Enhanced Populations}

	\begin{figure}[htp] %[t!]
	\centering
	\includegraphics[trim=0cm 0cm 0cm 0cm, scale=.38]{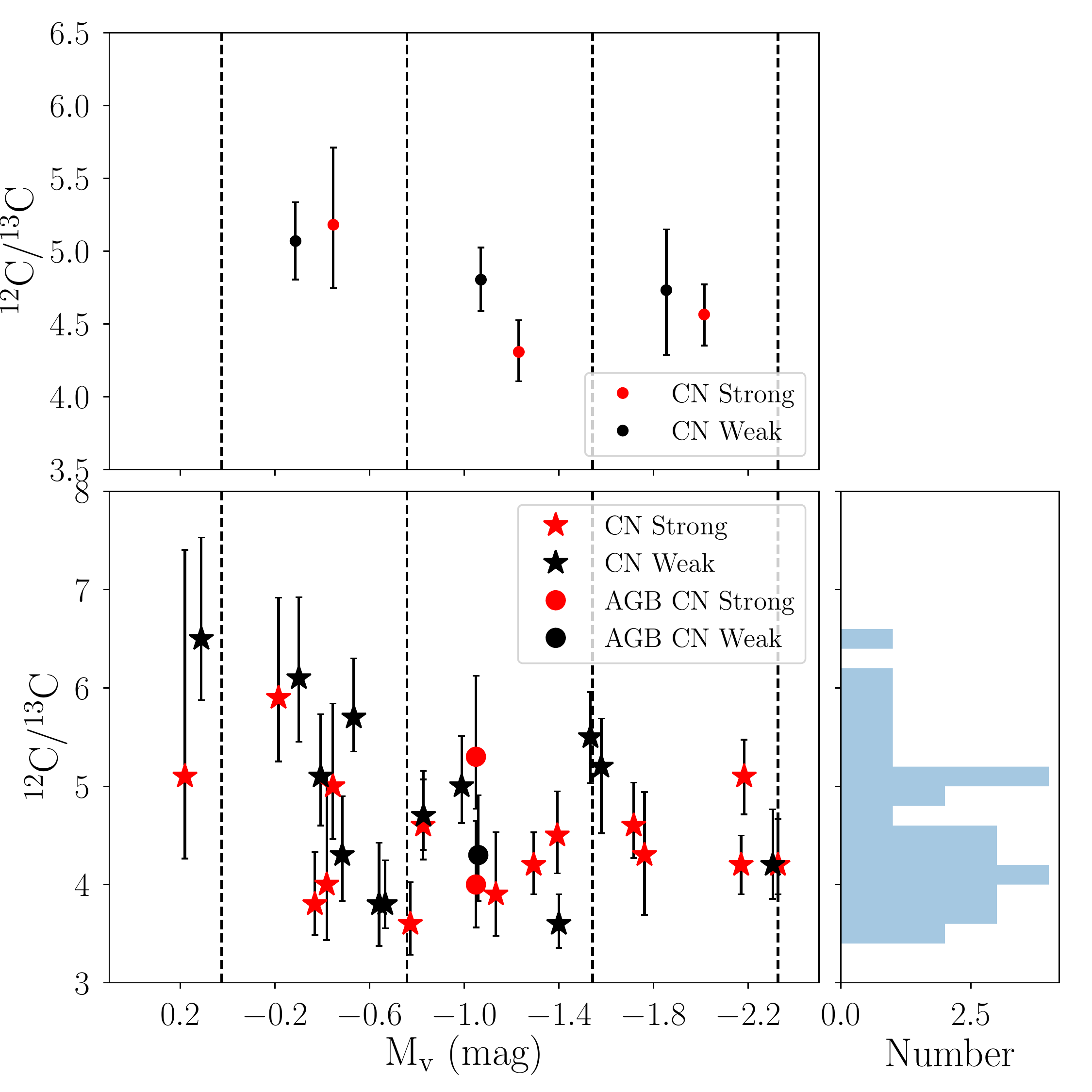}
	\caption{Bottom Left Panel: $\Cratio$ ratio versus the absolute V band magnitude for our sample stars. CN-enhanced stars are red, CN-normal stars are black in this plot. Stars on the red giant branch are represented by stars, AGB stars are circles. Dashed lines represent the location of the bin edges from Table \ref{table::binnedaverages}. Right: The right plot shows a histogram of $\Cratio$ measurements with a bin size of 0.2. Top: Plot shows the binned averages for the CN-enhanced and CN-normal red giant stars in each bin. \label{fig:binned_data}  }
	\end{figure}

Both the initial surface abundances of each star and evolutionary affects will determine the carbon isotope ratios of our sample. We can determine if any differences in carbon isotope ratio exist between the two populations of stars in M10 beyond the luminosity function bump where extra mixing is expected to bring $^{13}$C enriched material to the surface. 

Figure \ref{fig:binned_data} shows measured $\Cratio$ ratio as a function of absolute magnitude for our sample of 31 stars with CN-enhanced and CN-normal stars indicated as such. As expected based on other observations in GCs, the $\Cratio$ ratio decreases with luminosity for our faintest stars (M$_{\rm V} > -0.6$), but quickly reaches an asymptotic value and remains roughly constant for the brighter stars (M$_{\rm V} < -0.5$). However, there is no obvious difference between $\Cratio$ for CN-enhanced and CN-normal stars (i.e. two distinct tracks in Fig. \ref{fig:binned_data}). 

To determine if a small difference is hidden by uncertainties and not easily discernible by eye, more advanced analysis techniques were employed. We bin the data into four bins, choosing the bin size from Scott's rule\footnote{Scott's Rule: Number of bins equals $\frac{3.5 \sigma}{n^{(1/3)}}$ where n is number of data points}. We excluded the three AGB stars in our sample (IDs 67, 68, 69) and the two faintest stars with M$_{\rm V} > 0$ were placed in their own bin. We assume in each bin the evolutionary effects on surface abundances for each star are approximately the same for a given M$_{\rm V}$ and that the mean initial surface abundance is uniform for each population. Uniform initial surface abundances allow averaging the results in each bin, since abundance measurements in each star are treated as drawn from a population with the same mean and distribution.

Because the uncertainties in the $\Cratio$ ratios are asymmetric as represented by the error bars in Figure \ref{fig:binned_data}, a simple average of the data without properly considering the uncertainties may give incorrect or misleading results. To account for these asymmetric uncertainties, we generated random measurements with an uneven Gaussian distribution that matched the uncertainties. In order to properly quantify the uncertainties on any averages from the data, we adopt a likelihood function with asymmetric errors. We adopt the approximation of \citet{barlow04}. 

\begin{equation}
Log(L) = -\frac{1}{2} \frac{(\hat{x} - x)}{V + V'(x-\hat{x})} \ ,
\label{eq::likelihood_assymetric}
\end{equation}

\noindent where $x$ is the true value, $\hat{x}$ is the measured result, $V = \sigma_{-}\sigma_{+}$, and $V' = \sigma_{+} - \sigma_{-}$. The approximation is adopted because the true likelihood function is a mixture of uncertainties from the atmospheric parameters and model fit. We performed a Monte Carlo simulation based on the likelihood function to determine our average and uncertainties for each population in each bin and in the entire sample. In each bin, the summed likelihood function is estimated numerically by generating random observations using Equation~\ref{eq::likelihood_assymetric} and summing the results. The procedure was repeated for 100,000 sample sets of observations, giving a distribution of possible averages. The $50\, \%$ median value was taken as the summed correct value (or average), the $84\,\%$ value was taken as the upper 1$\sigma$ uncertainty, and the $14\, \%$ value was taken as the lower 1$\sigma$ uncertainty. The results from the Monte Carlo exercise for each bin are shown in Table \ref{table::binnedaverages} and the results are plotted in Fig. \ref{fig:binned_data}.

The binned data show no statistically significant differences between the CN-enhanced and CN-normal populations in any of the magnitude bins. Only in the --1.54 $<$ M$_{\rm V}$ $<$ --0.76 bin are the mean values of the CN-enhanced and CN-normal population different by $\sim$ 1-2 $\sigma$. We performed a t-test and Kolmogorov-Smirnov test for this bin and both returned p-values of 0.08 and 0.23. The null hypothesis that both sets of observations are drawn from the same underlying sample could not be rejected. We also calculated the two sided z score for the averages and mean standard deviation in this bin, for the number of stars (5 and 4) used to calculate the sample. We found a p-value of 0.38 from a Z score of 0.87, indicating we cannot rule out both samples having the same average carbon isotope ratio. Finally, we can also compare the average values for all the measured stars, excluding the three AGB stars, and the results are given in Table \ref{table::binnedaverages}.  The mean carbon isotope ratios for each complete sample are $\sim$ 1.2$\sigma$ - 1.5$\sigma$ apart, hinting at a difference between the populations but are not significant given the uncertainties on the $\Cratio$~ ratios.

Other studies of globular clusters find small to no differences in the carbon isotope ratio between different stellar populations, depending on metallicity. In M71 ($\FeH = -0.78$ \citealt{harris96} (2010 edition)), the CN-normal and enhanced populations showed average $\Cratio$~ abundances of 8.3 and 6, respectively \citep{briley94,briley97}. Further studies of M71 also found a slight offset in the carbon isotope between CN-normal and CN-enhanced populations \citep{smith07}. However, this effect has not been found in more metal poor globular clusters. For example, no difference in the carbon isotope ratios between CN-normal and CN-enhanced populations are found in M4 ($\FeH = -1.16$ \citealt{harris96}) and NGC 6752 ($\FeH = -1.54$ \citealt{harris96}) and all carbon isotope ratios are primarily between $\approx 4 \textrm{--} 6$ \citep{suntzeff91}. Finally, the carbon isotope ratio for stars brighter than the LFB in four clusters spanning $-2.45 < \FeH < -1.2$ are all near the CNO equilibrium value, between $\approx 4 \textrm{--} 10$ \citep{recio07}. While a difference in $\Cratio$ ratio may have existed in the two populations before evolving, the mixing of CNO process material is too efficient to detect any remaining differences in our evolved sample stars. 

\begin{deluxetable}{ c c c c c }
%\tabletypesize{10pt} 
%\rotate
\tablewidth{0pt} 
\tabletypesize{\footnotesize}
\tablecaption{Binned Carbon Isotope Ratio Averages\label{table::binnedaverages}} 
 \tablehead{\colhead{M$_{\rm V}$ Bin} & \colhead{$< \Cratio >$} & \colhead{$< \Cratio >$} & \colhead{N(Stars)} & \colhead{N(Stars)} \\
 \colhead{Width (mag)} & \colhead{CN-Enh.} & \colhead{CN-Norm.} &  \colhead{CN-Enh.} & \colhead{CN-Norm.}  }
\startdata
(--2.33 - --1.54)  &  4.56 $_{-0.20}^{+0.21}$ & 4.73 $_{-0.43}^{+0.45}$ & 5 & 2 \\
(--1.54 - --0.76)  & 4.31 $_{-0.20}^{+0.21}$ & 4.81 $_{-0.21}^{+0.22}$ & 5 & 4\\
(--0.76 -   0.02)  & 5.19 $_{-0.44}^{+0.52}$ & 5.08 $_{-0.25}^{+0.27}$ & 4 & 6\\
(0.02$<$)  &  5.2  & 6.5& 1 & 1 \\
All Stars\tablenotemark{a}  & 4.84 $_{-0.22}^{+0.27}$ & 5.10 $_{-0.17}^{+0.18}$ & 15 & 13 \\
\enddata
\tablenotetext{a}{Excludes AGB Stars (ID: 67, 68, 69)}
\end{deluxetable}

\section{Thermohaline Mixing Models: Model Parameters}
\label{sec::models1}
Stellar evolution models were created to explore the effects of extra mixing on the carbon isotope ratios, carbon abundances, and nitrogen abundances for stars in M10. Specifically, the models were created to determine if our $\Cratio$ measurements in stars near the luminosity function bump to the tip of the RGB provide a meaningful constraint to thermohaline mixing models and if so, do simple models accurately reproduce the multiple abundance trends measured in M10. Following previous studies (e.g. \citealt{angelou12,angelou15,henkel17}), we use the photometric and spectroscopic data available for M10 to model non-canonical mixing.

To build our simplified model, we use MESA (v. 10000: \citealt{paxton11,paxton13,paxton15,paxton18}) software suite\footnote{in-list found at \url{https://github.com/zm13}}. Previous studies have used MESA to construct stellar evolutionary tracks and isochrones; in particular we primarily adopt parameters from \citet{choi16} and the MIST project to build our thermohaline mixing models. Our model parameters are given in Table \ref{table::mesamodel}. However, for the purposes of this study, we make an important simplifying assumption. Since the RGB is a relatively short lived evolutionary phase, we assume the mass differences between the upper and lower RGB stars are negligible, especially concerning thermohaline mixing efficiency. From this assumption, we use a stellar evolutionary track for one representative star, instead of developing isochrones for our model. We also test our final stellar evolutionary track to representative MIST isochrones. Although differences will arise, especially between the base and tip of the giant branch, we can still accomplish our main science objective: to determine how well are the $\Cratio$, [C/Fe], and [N/Fe] trends predicted by the same thermohaline mixing model. 

\begin{deluxetable}{c c c c}
%\tabletypesize{10pt} 
%\rotate
\tablewidth{0pt} 
\tabletypesize{\small}
 \tablehead{\colhead{Parameters} & \colhead{Model Value} &  \colhead{Model Value} & \colhead{Parameter}  \\ \colhead{} & \colhead{(CN-Normal)} &  \colhead{(CN-Enhanced)} & \colhead{Reference}}
\tablecaption{Updated MESA Model Parameters  \label{table::mesamodel}} 
\startdata
Mass & 0.8 M$_{\odot}$ & \nodata \\
Y & 0.25  & 0.256, 0.28 & 1 \\
Z & 0.0015 & 0.0015 & \nodata \\
X($^{12}$C) & 7.16 x 10$^{-6}$ & 4.12 x 10$^{-6}$ & \nodata  \\
X($^{13}$C) & 7.36 x 10$^{-8}$ & 7.23 x 10$^{-8}$&\nodata \\
X($^{14}$N) & 5.24 x 10$^{-6}$ & 5.24 x 10$^{-5}$ &\nodata \\
X($^{16}$O) & 3.80 x 10$^{-5}$ & 1.174 x 10$^{-5}$ &\nodata \\
TO Age & 11.6 Gyr & 10.9 Gyr & \nodata \\
Opacity Mix. & [$\alpha$/Fe] = $+$0.4 & [$\alpha$/Fe] = $+$0.4 & 2 \\
$\alpha_{\rm MLT}$  & 2.05 & 2.05 & see text\\
$\alpha_{\rm Th}$  & 60, 666 & 60, 666 & 3 \\
$\eta_{RGB}$\tablenotemark{a} & 0.4 & 0.4 & 4\\
\enddata
\tablenotetext{a}{Mass loss scaling parameter}
\tablecomments{Sources: (1) \citealt{milone18}; (2) \citealt{grevesse98}; (3) \citealt{kippenhahn80}; (4) \citealt{reimers75}}
\end{deluxetable}

	\begin{figure}[htp]
	\centering
	\includegraphics[trim=.75cm 0cm 0cm 0cm, scale=.87]{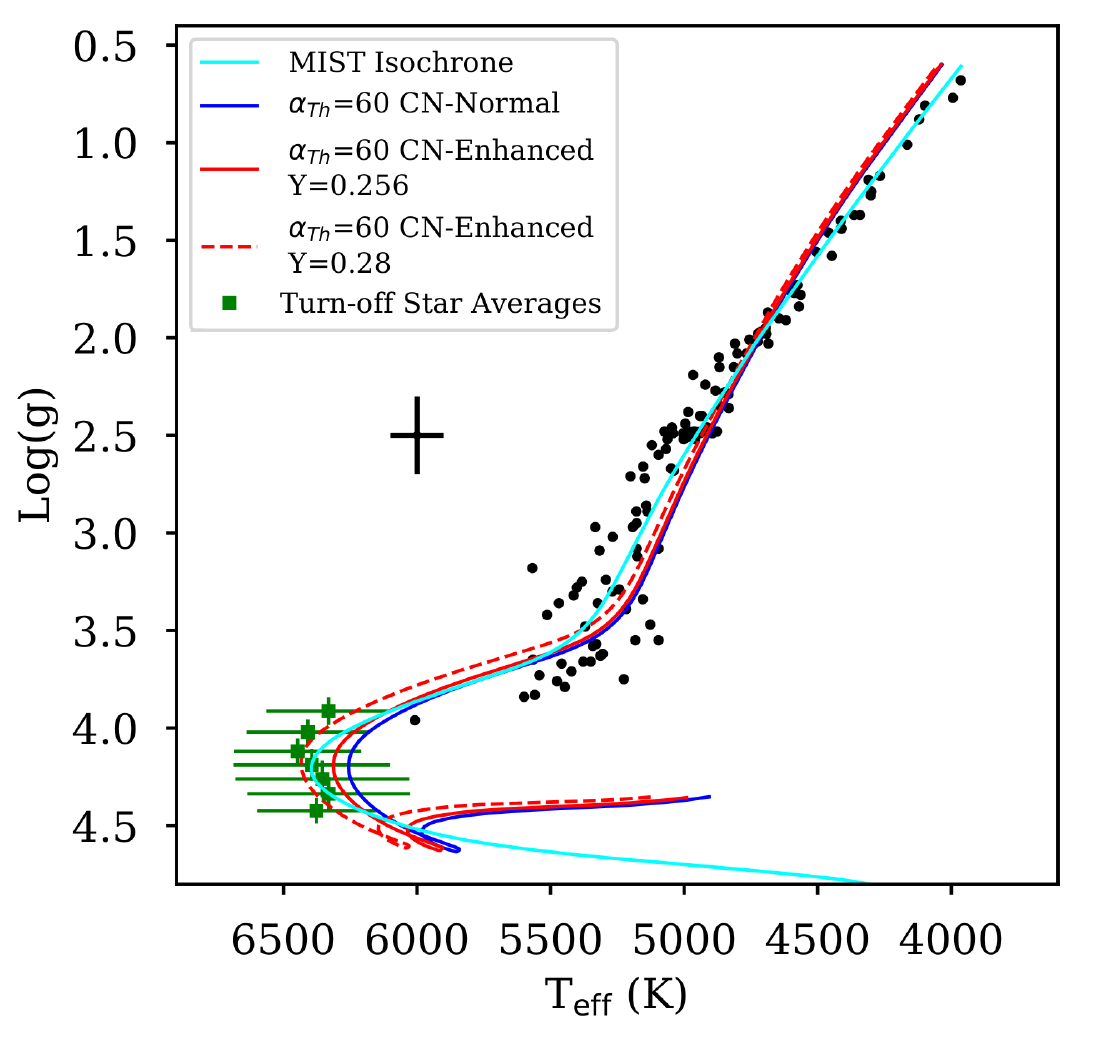}
	\caption{Temperature and log($g$) values for the entire sample of \citet{gerber18} compared to model predictions. Three models for the CN-normal, CN-enhanced with Y = 0.256, and CN-Enhanced with Y=0.28 are plotted. Cyan line is a MIST isochrone with [Fe/H] = --1.1 at an age of 11.7 Gyr.  A representative errorbar for the data is also plotted. Binned average atmospheric parameters for turn-off stars are shown as green squares (see sec. \ref{subsubsec::he_turnoff}). \label{fig:mesa_models_tefflog_best_models}  }
	\end{figure}

	\begin{figure*}[htp]
	\centering
	\includegraphics[trim=0cm 0cm 0cm 0cm, scale=.5]{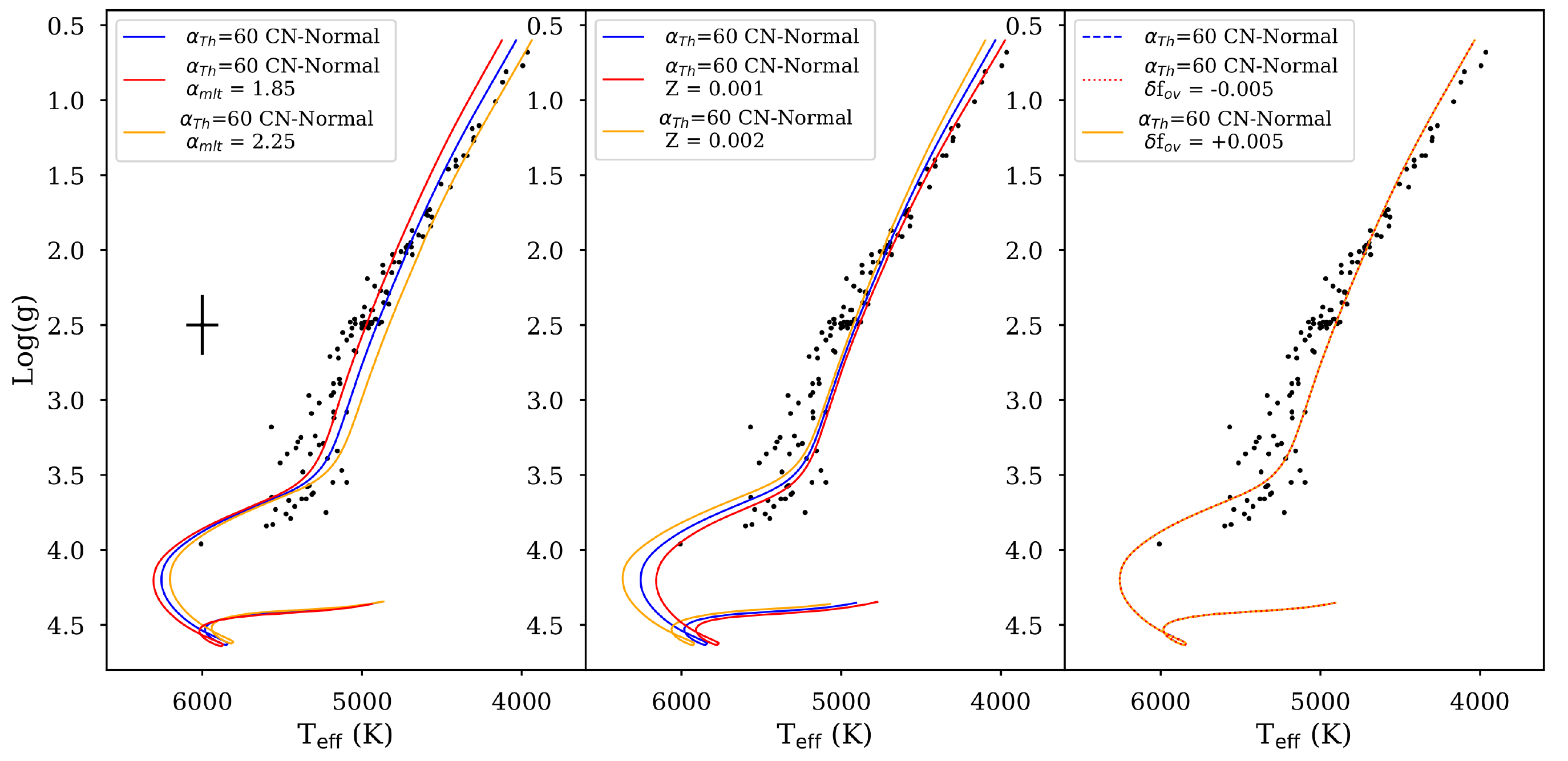}
	\caption{Temperature and log($g$) values for the entire sample from \citet{gerber18} compared to model predictions. Three models for the CN-Normal population are plotted in each window. Left: the mixing length $\alpha_{MLT}$ is varied by $\pm$ 0.2. Middle: Metallicity (Z) is varied by $\pm$ 0.0005. Right: the overshooting parameter (f$_{ov}$) is varied by $\pm$ 0.005. A representative errorbar for the data is also plotted. \label{fig:mesa_models_tefflog}  }
	\end{figure*}

\subsection{He Abundance}
\label{subsubsec::he_turnoff}
We set the He abundance based on cosmic abundances and He estimations in M10 \citep{milone18}. The CN-normal population model uses Y = 0.25, consistent with the primordial He abundance. Two separate models were created for the CN-enhanced population to represent the spread of possible helium abundances for second generation stars. The first model was created with Y = 0.256, using the $\Delta$Y spread between the first and second generation in M10 derived from \citet{milone18}. The second CN-enhanced model was made with Y=0.28, corresponding to the maximum change in Y possible for the second generation of stars in M10 \citep{milone18}.

We tested to see if the stellar models matched the turn-off region in M10 using photometry from \citet{pollard05}. The turn-off region is too faint for 2MASS (K$_{s}$ $<$ 14.3 limit) and so we used V and B-V photometry \citet{pollard05} and the T$_{\mathrm{eff}}$-color relation of \citet{casagrande06}, which unlike \citet{alonso99} is calibrated to dwarf stars. Uncertainties in the effective temperatures of the turn-off stars are large due to the scatter in the effective temperature relation and spread in the V magnitudes at the turn off (e.g. Fig. 1 from \citealt{gerber18}). To better constrain the models, we computed average T$_{\mathrm{eff}}$ and log(g) values in M$_{v}$ bins of size 0.2 mag with a range from 3.8 $<$ M$_{v}$ $<$ 5.2; examining only turn-off stars. The average parameters and the standard deviations for each bin are shown in Fig. \ref{fig:mesa_models_tefflog_best_models}. We demonstrate that our models are consistent within the 1$\sigma$ uncertainties with the average turn-off stars although the MIST isochrones and Y=0.28 models most closely match the photometric atmospheric parameters. 

\subsection{Convective Overshooting}

We adopted the exponential overshooting of \citet{herwig00} which has two parameters controlling the efficiency of convective overshoot: $f$ and $f_{0}$. In this formalism, $f$ is the efficiency of the convective overshooting and $f_{0}$ is used to define the depth from the boundary in which convective overshooting begins. We match the values from \citet{choi16} of $f_{ov,core}=0.016$ and  $f_{ov,envelope}=0.0174$, and set $f_{0} = f/2$. Our overshooting parameter causes a reasonable evolution towards the turn-off, as shown in the right panel of Figure \ref{fig:mesa_models_tefflog}. 

To further test our over-shooting parameter, we created models for the CN-normal population with the overshooting parameter ($f$) changed by $\pm$ 0.005. We note the higher end of this parameter space explores the maximum value for the convective overshooting parameter ($f$ = 0.02) for exponential overshooting used in other works \citep{magic10}. We find no significant changes for this range of overshooting parameters on our atmospheric parameters (Fig. \ref{fig:mesa_models_tefflog}). Finally, we tested the effects of changing the overshooting parameter on the extra mixing beyond the luminosity function bump. We found no significant deviations from our best fit model as shown in Fig. \ref{fig:mesa_models_abun}.  

\subsection{Metallicity}

We adjusted the metallicity of our stars and the mixing length parameters ($\alpha_{MLT}$) to better model the atmospheric parameters of our sample of stars. We found a metallicity of Z = 0.0015 created a model that reasonably fits the observations. The \citet{harris96} (v.2010) metallicity for M10 is given as [Fe/H] = --1.56; Z = 0.0015 can be achieved for a composition mixture of \citet{grevesse98} with an [$\alpha$/Fe] = 0.57 enhancement, and re-normalizing the [Fe/H] abundance in the Harris catalog  \citep{gratton03, carretta09b} to the scale of \citet{grevesse98}. The average [O/Fe] = 0.41 $\pm$ 0.15, [Mg/Fe] = 0.49 $\pm$ 0.04, and [Si/Fe] = 0.28 $\pm$ 0.05 \citep{carretta09b}, making our assumption high but not unreasonable. We note photometric observations of M10 place the location of the LFB at M$_{\mathrm{V}}$ $\sim$ 0.7 \citep{nataf13}. In order to match the location of the luminosity function bump, our models were shifted by $+0.40\, \mathrm{mags}$, comparable to shifts found for other clusters with similar metallicities \citep{zoccali00,meissner06,cassini11,angelou15}.

\subsection{Mixing Length Parameter}

The mixing length parameter is significantly less well defined for M10 giants than the other atmospheric parameters. Other studies of metal poor clusters have adopted the solar mixing length parameter to model their clusters \citep{angelou15}. We varied the mixing length parameter, using the mixing length theory of \citet{henyey65}, and found higher values are needed to fit the most evolved red giants while smaller values are needed for the base of the red giant branch, as shown in Fig. \ref{fig:mesa_models_tefflog_best_models}. The most reasonable value that most closely fits the derived atmospheric parameters is $\alpha_{\rm MLT}$ = 2.05. We varied the mixing length parameter by 0.2, demonstrating these effects in the left panels of Fig. \ref{fig:mesa_models_tefflog},\ref{fig:mesa_models_abun}. 

In other studies the mixing length parameter has been found to vary with stellar evolution and metallicity. A study of metal rich red giants from Kepler has found $\alpha_{\rm MLT}$ between $\approx 2 \textrm{--} 2.2$ \citep{li18}. Other studies of Kepler stars find the mixing length parameter is metallicity dependent; as the metallicity decreases the mixing length also decreases \citep{tayar17}. 

\subsection{Additional Model Parameters}

We used the opacity mixture of gs98$\_$aFe$\_$p4 in MESA, which use scaled solar abundances from \citet{grevesse98} with $\alpha$ elements enhanced by 0.4 dex to match the $\alpha$ enhancement measured by \citet{carretta09b}. The \mesa \ reaction net that included CNO reactions\footnote{specifically \texttt{cno\_extras\_o18\_to\_mg26\_plus\_fe56}} was adopted to include all CNO related nucleosynthesis processes occurring in the red giant stars. The mass fractions for the relevant CNO elements are given in Table~\ref{table::mesamodel}. The $\alpha$ elements in this network, oxygen, neon, and magnesium were assumed to be $0.39\, \mathrm{dex}$ larger than the scale with [Fe/H], adopted from the average abundance for the alpha elements in \citet{carretta09b}. Other mass fractions in the net are scaled to solar abundances adopted from \citet{asplund09} for $\FeH =-1.56$. 

We also used a Reimers mass loss parameter of 0.4; adopted based on similar studies creating stellar evolution tracks and isochrones for low mass stars in the literature \citep{girardi00,pietrinferni04,ekstrom12}. The turn-off age given in Table \ref{table::mesamodel} is defined as the temperature inflection point; when main sequence stars begin to cool and fall onto the subgiant branch. Using these definitions we find a turn-off age of 11.6 Gyr for the CN-normal stars, and an age of 10.9 Gyr for the CN-enhanced stars. Our CN-normal age is consistent with a cluster age of 11.75 $\pm$ 0.38 Gyr derived using cluster photometry \citep{vandenberg13}.

The cause of the age inconsistency is due to approximating the red giant branch evolution with a stellar evolutionary track; both small changes in the mass and He abundance will determine the age of the star during the turn-off. Since the second generation of stars will have an age difference on the order of $>$0.1-0.3 Gyr, the masses of second generation and first generation stars at identical points on the CMD will be slightly different. To test the significance of this effect, we ran a second generation model but changed the mass to 0.79 M$_{\odot}$ and determined a turn off age of 11.5 Gyr. Thermohaline mixing models with a difference of 0.01 M$_{\odot}$ are nearly identical to the 0.8 M$_{\odot}$ models and cannot explain the difference between the $\Cratio$ model fit and the [C/Fe] ratio fits. Additionally, the MIST isochrone is composed of masses ranging from 0.78 M$_{\odot}$ to 0.83 M$_{\odot}$ from the turn off to main sequence and the thermohaline mixing models from the isochrone are consistent with our stellar evolutionary track. We therefore use 0.8 M$_{\odot}$ as a representative mass for the giant branch models. 

Finally, we note there are systematic differences between the model and the observed parameters as shown in Fig. \ref{fig:mesa_models_tefflog_best_models}. First, the systematic differences may be due to using a stellar evolution model instead of an isochrone. For example, not treating the small mass differences from the tip of the RGB will affect the model. To test this, we compare our tracks with a MIST isochrone (v. 1.2), created with at an age of 11.7 Gyr and an [Fe/H] = -1.1. The MIST isochrones are more limited, the [Fe/H] of -1.1 matches our metallicity (Z) for \citet{asplund09} scaled abundances used in MIST. 

From Figure \ref{fig:mesa_models_tefflog_best_models} we find the isochrones provide a better fit to the CMD than the single stellar track, especially at the tip and base of the RGB, and match the model best for stars near the luminosity function bump. We also note a mixing length parameter that varies as a function of evolution (lower for the base of the RGB, higher for the tip), would fit the cluster best as demonstrated in Fig. \ref{fig:mesa_models_tefflog}. Next, there may be systematic uncertainties in atmospheric parameters for our coolest stars, where the color-T$_{\mathrm{eff}}$ determination becomes more uncertain. However, the agreement between the model and our atmospheric parameters is sufficient for our analysis and reasonable changes in parameters that may affect the atmospheric parameters do not significantly alter our conclusions about the $\Cratio$. 

\section{Thermohaline Mixing Models: Results}%: Constraining $\alpha_{\mathrm{TH}}$}
\label{sec::models2}
\subsection{Constraining $\alpha_{\mathrm{TH}}$}
The thermohaline mixing diffusion coefficient ($\alpha_{\rm Th}$) is approximated in $\mesa$ \citep{paxton13} and used to scale the efficiency of thermohaline mixing. ($\alpha_{\rm Th}$) was adjusted to best fit the carbon, nitrogen, and carbon isotope ratios measured in M10 stars. We adopted the thermohaline mixing theory of \citet{kippenhahn80} for our model. We attempted to match the abundance patterns of carbon, nitrogen, and the carbon isotope ratio by adjusting the thermohaline coefficient, with models shown in Figure \ref{fig:mesa_models_abun_best}. When fitting the abundances we first attempted to only fit the first, `primordial' generation of stars observed in M10 (CN-normal) to determine the efficiency needed. We then changed our initial abundances to those found in \citet{gerber18} for the second generation (CN-enhanced) (i.e. [C/Fe]~$\sim$~--0.25, [N/Fe]~$\sim$~1.3, [O/Fe]~$\sim$~0.11) listed in Table \ref{table::mesamodel}, to determine a rate for both populations. We note the initial [N/Fe] abundance for the CN-normal population was set to approximately the average [N/Fe] abundance at the luminosity function bump. Abundances beyond M$_{v}$ $>$ 2 had significantly larger error-bars and measurements near the function bump give more accurate representation of the abundances. Section 3.2 discusses the uncertainties in \citet{gerber18} and additionally Fig. 10 in \citet{gerber18} shows the locus at [N/Fe] $\sim$ 0.5 for the CN-normal population.

The carbon and nitrogen abundances of both populations are fit best with an $\alpha_{\rm Th}$ $\sim$ 60 as seen in Figure \ref{fig:mesa_models_abun_best}. This result matches what was observed by \citet{gerber18} who found that both populations were depleting in carbon at the same rate. We also plot the surface abundances from the MIST isochrone in Fig. \ref{fig:mesa_models_abun_best}, which is set with an $\alpha_{\rm Th}$ of 666. The results from the isochrone are consistent with the results of the evolutionary tracks. Further MIST isochrones were not computed due to the limited number of input parameters available on the MIST web interpolator. Only the [Fe/H] abundance and age were easily changeable. By modelling our own stellar evolution tracks we may change the initial abundances to match M10 and most importantly vary the $\alpha_{\rm Th}$ parameter. The MIST isochrone abundances and M$_{\mathrm{v}}$ have been shifted to match the input M10 abundances and luminosity function bump location.

\begin{figure*}[htp]
\centering
\includegraphics[trim=1cm 0cm 0cm 0cm, scale=.73]{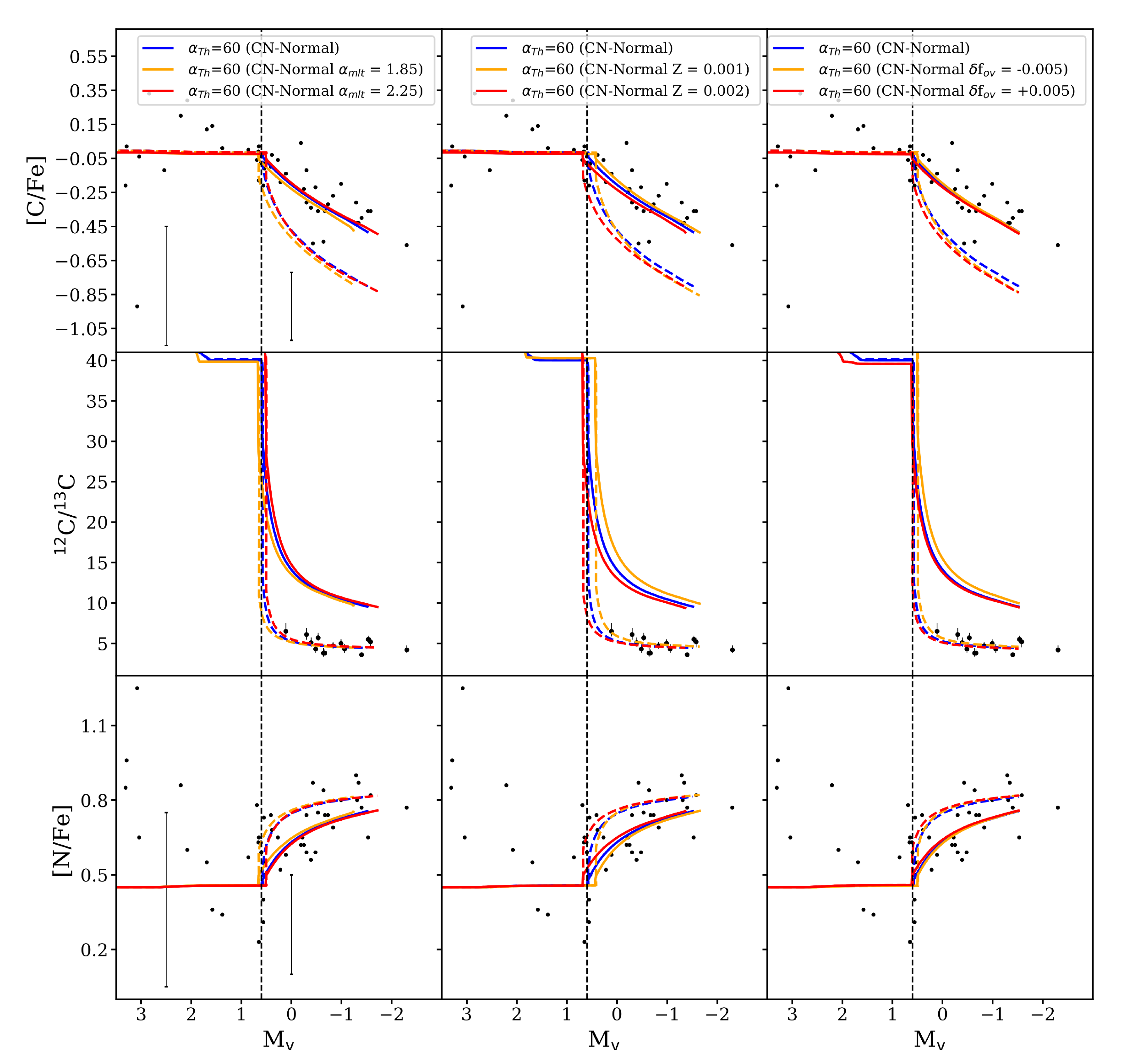}
\caption{$\Cratio$ ratios from this work with carbon and nitrogen abundances from \citet{gerber18} for the CN-normal population compared to model predictions. Left: the mixing length $\alpha_{MLT}$ is varied by $\pm$ 0.2. Middle: Metallicity (Z) is varied by $\pm$ 0.0005. Right: the overshooting parameter (f$_{ov}$) is varied by $\pm$ 0.005. Dashed line indicates the location of the luminosity function bump. Solid lines represent models with $\alpha_{TH}$ = 60 while dashed lines have $\alpha_{TH}$ = 666. Representative errorbars for the carbon and nitrogen abundances measured in \citet{gerber18} are plotted for two magnitudes in the left panels.
\label{fig:mesa_models_abun} }
\end{figure*}

\begin{figure}[htp]
\centering
\includegraphics[trim=0cm 0cm 0cm 0cm, scale=.64, clip=True]{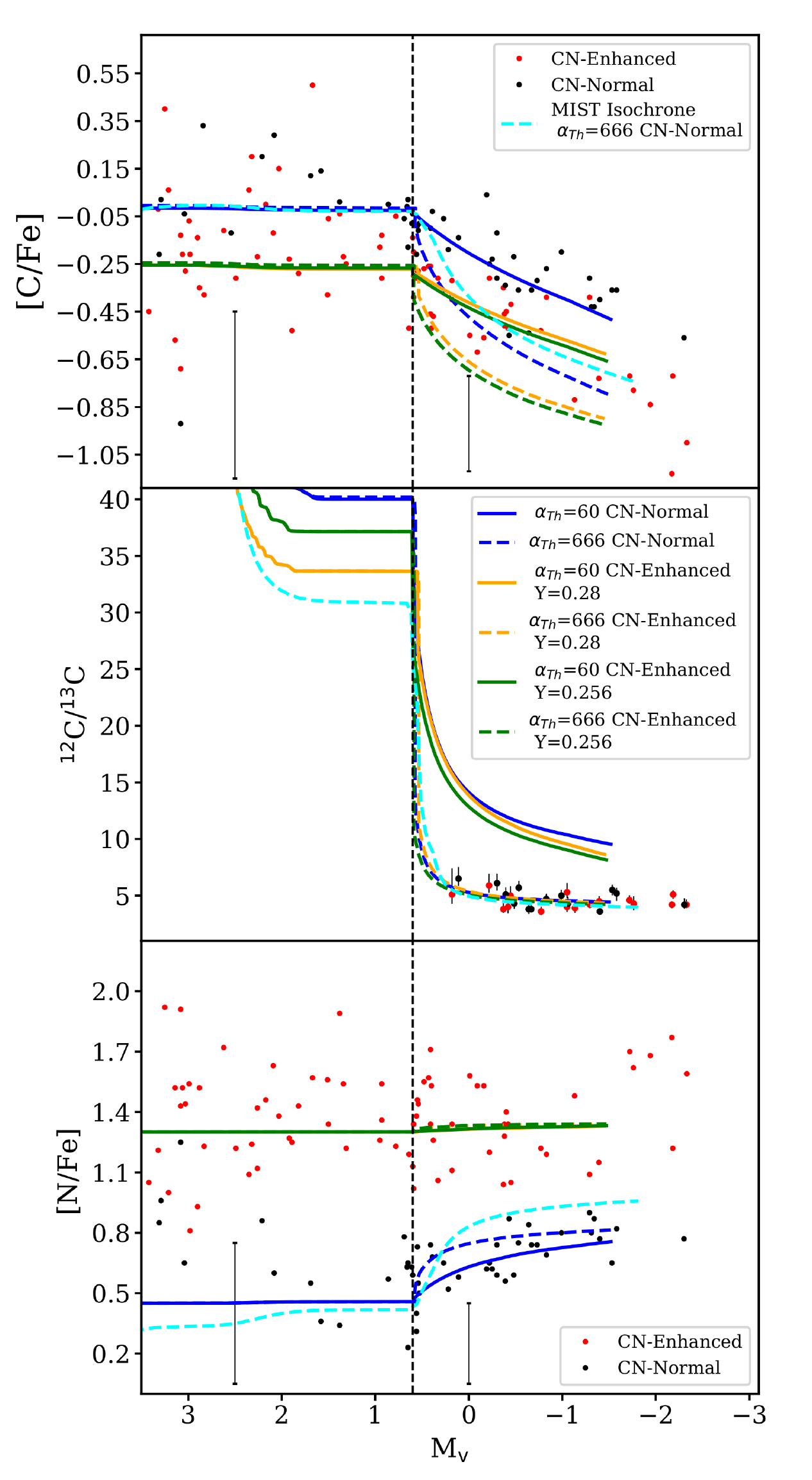}
\caption{$\Cratio$ ratios from this work with carbon and nitrogen abundances from \citet{gerber18} compared to model predictions. The blue lines are for models matching the first generation, green lines for the second generation with Y=0.256,and the orange lines are for models matching the second generation with Y = 0.28. Cyan line is a MIST isochrone with [Fe/H] = --1.1 at an age of 11.7 Gyr. Models with $\alpha_{\rm Th} = 60$ are shown as solid lines and models with $\alpha_{\rm Th} = 666$ are shown as dashed lines. The black points represent CN-normal stars and the red data points represent CN-enhanced stars. The top panel compares carbon abundances, the middle the $\Cratio$ ratio, and the bottom compares nitrogen abundances to the absolute V-band magnitude (M$_{\mathrm{v}}$) of each star. Dashed line indicates the location of the luminosity function bump. Representative errorbars for the carbon and nitrogen abundances measured in \citet{gerber18} are plotted for two magnitudes.
 \label{fig:mesa_models_abun_best} }
\end{figure}

\subsection{Discussion on Model Fits}

Once we were able to match the observed carbon and nitrogen abundances with a model, we then attempted to match the carbon isotope ratios with similar models. We find two different thermohaline efficiencies are necessary to fit both the carbon abundances and the carbon isotope ratios; the $\alpha_{\rm Th}$ must be enhanced from 60 to 666 (a typical $\alpha_{TH}$ value representing the upper range of thermohaline mixing used in the literature \citealt{paxton13,choi16}) to match the low $\Cratio$~ratios found in M10. Other studies also find that the efficiency of mixing is two orders of magnitude lower than needed to match surface abundance in stars compared to predicted $\alpha_{TH}$ = 1 \citep{wachlin14}. Also different thermohaline efficiencies are needed to match both lithium and carbon abundances in globular cluster giants \citep{angelou15,henkel17}. The need for different efficiencies to match different observables affected by mixing indicates that the prescription of thermohaline mixing may need to be modified (as was done by \citealt{henkel17}) to create a prescription that matches carbon and lithium abundances. We therefore have shown that the $\Cratio$ adds an important additional constraint to any effort to model all surface abundances affected by non-canonical mixing simultaneously.

Finally, we tested whether changes in our model parameters could explain the discrepancy in $\alpha_{TH}$ needed to fit the carbon abundances and $\Cratio$ ratios in our stars. In Fig. \ref{fig:mesa_models_abun} we show the $\Cratio$ ratio cannot be reproduced with our simple model and more advanced modeling or adjustments to the thermohaline mixing theory are necessary to match our observations. We found changes in the mixing length parameter over a range of $\pm$ 0.2 do not affect the main conclusion that different thermohaline mixing efficiencies are needed to explain the decline of the $\Cratio$ and [C/Fe]. Simnilarly, chages in Z by $\pm$ 0.0005 and the overshooting parameter of $\pm$ 0.005 were not able to match the range of thermohaline mixing efficiencies needed to match the abundance trends.

\section{Summary}
\label{sec::conclusion}

We measured carbon abundances and $\Cratio$ ratios in 31 stars spanning -2.33 $<$ M$_{\rm V}$ $<$ 0.18 in M10 using CO lines at $\sim 2.3 \textrm{--} 2.4\, \mu \mathrm{m}$. Synthetic spectra were generated using MOOG \citep{sneden73}, MARCS model atmospheres \citep{gustafsson08}, and atmospheric parameters from \citet{gerber18}. Carbon abundances and $\Cratio$ ratios were derived by finding the synthetic spectrum that minimized the $\chi^{2}$ between the generated spectrum and observations for each star. These abundance measurements yield multiple results:

\begin{enumerate}

\item{Carbon abundances using the infrared CO lines agree with measurements using CH and CN features from \citet{gerber18}; the average difference between the [C/Fe] ratios is --0.06 $\pm$ 0.16 (st.dev.) dex. The constancy between results suggests model 3D effects on CO lines at $\FeH = -1.56$ are similar to those on CN and CH.}

\item{The average carbon isotope ratio for the first generation (CN-normal) red giant branch stars in M10 is $\Cratio$ = 5.10$_{-0.17}^{+0.18}$ (13 stars total) and for the second generation (CN-enhanced) $\Cratio$ = 4.84$_{-0.22}^{+0.27}$ (15 stars). We found no statistically significant difference in the carbon isotope ratio between the two stellar populations. We also binned the data to group stars at similar stages of stellar evolution, and therefore dredged-up material, along the giant branch. In our bin of --1.54 $<$ M$_{\mathrm{v}}$ $<$ --0.76, there was a difference of $\sim$ 1-2 $\sigma$, also hinting at a difference between the two populations. We however found no difference within the errors for the other two bins.} The extra-mixing process is more efficient at lower metallicities and likely the non-canonical extra mixing has removed any potential initial differences in $\Cratio$ for the stars in M10. We also note all of our stars are above the luminosity function bump of M10 (M$_{\rm V}\sim 0.7\, \mathrm{mag}$; \citealt{nataf13}).

\item{We modeled the evolution of carbon, nitrogen, and the $\Cratio$ ratio for M10 red giants using \mesa. A thermohaline mixing efficiency of 60} ($\alpha_{\rm Th}$ = 60) \citep{kippenhahn80} to match the depletion of carbon in the surface of the star and by a factor of 666 ($\alpha_{\rm Th}$ = 666) to match the observed $\Cratio$ ratios.

\item{Two different thermohaline mixing efficiencies required to match the rates of decline of the carbon abundance and the $\Cratio$ ratio suggest more complex mixing formulations are required to match abundances. Observations of the carbon isotope ratios at a large range of M$_{\rm V}$ provide an additional constraint on mixing models. In conjunction with other abundances such as Li, C, the $\Cratio$ ratio, and N, may help decide between possible mixing models.}

\end{enumerate}

\acknowledgements
This work is based on observations obtained at the Gemini Observatory, which is operated by the Association of Universities for Research in Astronomy, Inc., under a cooperative agreement with the NSF on behalf of the Gemini partnership: the National Science Foundation (United States), the National Research Council (Canada), CONICYT (Chile), Ministerio de Ciencia, Tecnolog\'{i}a e Innovaci\'{o}n Productiva (Argentina), and Minist\'{e}rio da Ci\^{e}ncia, Tecnologia e Inova\c{c}\~{a}o (Brazil). The Gemini observations were done under proposal ID GN-2017A-Q-70. We thank Andr\'{e}-Nicolas Chen\'{e} for his assistance with the Gemini South Telescope observing run. This research has made use of the NASA Astrophysics Data System Bibliographic Services, the Kurucz atomic line database operated by the Center for Astrophysics. This research has made use of the SIMBAD database, operated at CDS, Strasbourg, France. This publication makes use of data products from the Two Micron All Sky Survey, which is a joint project of the University of Massachusetts and the Infrared Processing and Analysis Center/California Institute of Technology, funded by the National Aeronautics and Space Administration and the National Science Foundation. We thank Eric Ost for implementing the model atmosphere interpolation code. We thank the anonymous referee for their thoughtful comments and suggestions on the manuscript. C. A. P. acknowledges the generosity of the Kirkwood Research Fund at Indiana University. Z. G. M. acknowledges the Indiana University College of Arts and Sciences for research support via a Dissertation Research Fellowship.

\software{IRAF, MOOG (v2017; \citealt{sneden73}), MESA (v10000; \citealt{paxton11,paxton13,paxton15,paxton18}), \texttt{scipy} \citep{jones01}, \texttt{numpy} \citep{walt11}, \texttt{matplotlib} \citep{hunter07}, \texttt{emcee} \citep{foreman13}}

%\FloatBarrier

\FloatBarrier

\end{document}